\documentstyle[aps,12pt]{revtex}

\begin{document}
\author{Giuseppe Castagnoli\thanks{%
Information Technology Dept., Elsag Bailey, 16154 Genova, Italy}, Dalida
Monti\thanks{%
Universit\`{a} di Genova and Elsag Bailey, 16154 Genova, Italy}}
\title{A diakoptic approach to quantum computation}
\maketitle
\date{{}}

\begin{abstract}
In the diakoptic approach, mechanisms are divided into simpler parts
interconnected in some standard way (say by a ``mechanical connection''). We
explore the possibility of applying this approach to quantum mechanisms: the
specialties of the quantum domain seem to yield a richer result. First parts
are made independent of each other by assuming that connections are removed.
The overall state would thus become a superposition of tensor products of
the eigenstates of the independent parts. Connections are restored by
projecting off all the tensor products which violate them. This would be
performed by particle statistics, under a special interpretation thereof.
The NP-complete problem of testing the satisfiability of a Boolean network
is approached in this way. The diakoptic approach appears to be able of
taming the quantum whole without clipping its richness.

\end{abstract}

\date{}

\section{Definition of quantum mechanical connection}

In (classical) applied mechanics, the diakoptic (dissectionistic) approach
is exemplified by the notion of mechanical Connection. Connections divide
the whole into simpler parts and reconstruct it $-$ they introduce a
``divide and conquer'' strategy. In fig. 1(a), a crank-shaft is the
Connection which imposes an invertible function between the positions of
parts $r$ and $s$ (discretized as 0 and 1, then the function is the Boolean
NOT).

Things can be more difficult in quantum mechanics, since the Connection
Hamiltonian may not commute with the parts Hamiltonians. This difficulty is
avoided by implementing each Connection through a form of constructive and
destructive interference, assumedly related to particle statistics. By
applying reverse engineering, the Connection is first introduced as a
mathematical feature that would be nice-to-have in quantum mechanisms. Then
we ask ourselves whether that feature can be physical.

Let us consider the mechanism of fig. 1 from a quantum mechanical
perspective. The Connection should establish a constraint between two
otherwise independent quantum parts $r$ and $s$, with eigenstates $\left|
0\right\rangle _{r}$, $\left| 1\right\rangle _{r}$ and $\left|
0\right\rangle _{s}$, $\left| 1\right\rangle _{s}$ (fig. 1b). The Connection
state should have the form

\[
\left| \varphi \right\rangle =\alpha \left| 0\right\rangle _{r}\left|
1\right\rangle _{s}+\beta \left| 1\right\rangle _{r}\left| 0\right\rangle
_{s}\text{, with }\left| \alpha \right| ^{2}+\left| \beta \right| ^{2}=1. 
\]
We should note that the eigenvalues of each tensor product satisfy the
Boolean NOT $-$ the constraint; $\left| \varphi \right\rangle $ is free to
``move'' in a two-dimensional Hilbert space: this gives the one degree of
freedom required from a Connection. 

Let us assume that the Connection is temporarily removed. The generic state
of the independent parts are $\left| \Psi \right\rangle _{r}=\alpha
_{r}\left| 0\right\rangle _{r}+\beta _{r}\left| 1\right\rangle _{r},\ \left|
\Psi \right\rangle _{s}=\alpha _{s}\left| 0\right\rangle _{s}+\beta
_{s}\left| 1\right\rangle _{s}$. The whole unentangled state in the Hilbert
space of the two qubits $H_{w}$ is

\[
\left| \Psi \right\rangle =\alpha _{0}\left| 0\right\rangle _{r}\left|
0\right\rangle _{s}+\alpha _{1}\left| 0\right\rangle _{r}\left|
1\right\rangle _{s}+\alpha _{2}\left| 1\right\rangle _{r}\left|
0\right\rangle _{s}+\alpha _{3}\left| 1\right\rangle _{r}\left|
1\right\rangle _{s}, 
\]
with $\alpha _{o}=\alpha _{r}\alpha _{s}$, etc. The Connection is restored
by projecting $\left| \Psi \right\rangle $ on the ``symmetric'' subspace $%
H_{s}=span\left\{ \left| 0\right\rangle _{r}\left| 1\right\rangle
_{s},\left| 1\right\rangle _{r}\left| 0\right\rangle _{s}\right\} .$ Let us
define the projector (or ``symmetry'') $A_{rs\text{ }}$by:

\begin{eqnarray*}
A_{rs}\left| 0\right\rangle _{r}\left| 1\right\rangle _{s} &=&\left|
0\right\rangle _{r}\left| 1\right\rangle _{s},\ A_{rs}\left| 1\right\rangle
_{r}\left| 0\right\rangle _{s}=\left| 1\right\rangle _{r}\left|
0\right\rangle _{s}, \\
A_{rs}\left| 0\right\rangle _{r}\left| 0\right\rangle _{s} &=&\ A_{rs}\left|
1\right\rangle _{r}\left| 1\right\rangle _{s}=0.
\end{eqnarray*}

The $A_{rs}$ projection of $\left| \Psi \right\rangle $ is the normalized
vector of $H_{s}$ closest to it. This is obtained (in a peculiar way whose
motivation will be clarified) by submitting a {\em free normalized vector} $%
\left| \varphi \right\rangle $ of $H_{w}$ (whose amplitudes on the basis
vectors of $H_{w}$ are free and independent variables up to normalization)
to the mathematically simultaneous conditions: (i) $A_{rs}\left| \varphi
\right\rangle =\left| \varphi \right\rangle ,$ and (ii) the distance between
the vector before projection $\left| \Psi \right\rangle $ and that after
projection $\left| \varphi \right\rangle $ should be minimum; in equivalent
terms $\left\| \left\langle \Psi \right. \left| \varphi \right\rangle
\right\| $ should be maximum. This yields the usual result $\left| \varphi
\right\rangle =k\left( \alpha _{1}\left| 0\right\rangle _{r}\left|
1\right\rangle _{s}+\alpha _{2}\left| 1\right\rangle _{r}\left|
0\right\rangle _{s}\right) ,$ where $k$ is a renormalization factor, namely
an allowed Connection state. The Connection will perform by operating on the 
{\em parts} under continuous $A_{rs}$ projection on $H_{s}$ of the {\em whole%
}. This will turn out to be the nice-to-have mathematical feature.

\section{A diakoptic interpretation of particle statistics}

To give an introductory example, let us introduce a sort of Connection
simply related to particle statistics. Let 1 and 2 be two free, identical
and non-interacting spin 1/2 particles. At a given time, their overall
spatial wave function is a symmetrical/antisymmetrical linear combination of
the spatial wave functions of the two free particles, ($x_{1\text{ }}$and $%
x_{2}$ are the particles spatial coordinates):

\[
\Psi \left( x_{1,}x_{2}\right) =e^{ik_{A}x_{1}}e^{ik_{B}x_{2}}\pm
e^{ik_{A}x_{2}}e^{ik_{B}x_{1}}, 
\]
the $+$ ($-$) sign goes with the spin singlet (triplet) state (normalization
is disregarded). It can be seen that $\left\| \Psi \left( x_{1,}x_{2}\right)
\right\| ^{2}=\cos ^{2}kx\ $for the singlet state, and $\left\| \Psi \left(
x_{1,}x_{2}\right) \right\| ^{2}=\sin ^{2}kx\ $for the triplet state, where $%
x=x_{1}-x_{2},\ k=k_{A}-k_{B}$. Thus close (separated) particles are more
likely to be found in a singlet (triplet) state. There is a sort of
Connection inducing a correlation between the mutual distance of the two
particles and the character of their spin state: in principle, by operating
on the distance, the character of the spin state is (probabilistically)
changed. Noticeably, this kind of Connection would fall apart if the two
particles were not identical.

The Connection of Section I is a different case. It still relies on particle
statistics, but under a special interpretation thereof. A particle
statistics symmetry should be seen as the result of continuous projection of
the system state on a given symmetrical subspace. This interpretation can be
exemplified by considering a pair of identical bosons labeled 1 and 2; $%
S_{12}=\frac{1}{2}\left( 1+P_{12}\right) $ is the usual symmetrization
projector; 0/1 stand for, say, horizontal/vertical polarization. The
symmetry $S_{12}\left| \Psi \right\rangle =\left| \Psi \right\rangle $ is
satisfied in 
\[
H_{t}=span\left\{ \left| 0\right\rangle _{1}\left| 0\right\rangle
_{2},\left| 1\right\rangle _{1}\left| 1\right\rangle _{2},\frac{1}{\sqrt{2}}
\left( \left| 0\right\rangle _{1}\left| 1\right\rangle _{2}+\left|
1\right\rangle _{1}\left| 0\right\rangle _{2}\right) \right\} .
\]

There is a common didactic way of introducing this kind of symmetry. First,
statistics is disregarded and the particles are assumed to be independent of
each other. Let their {\em unentangled} state at some given time be $\left|
\Psi \left( t\right) \right\rangle ^{^{\prime }}=\alpha _{0}\left|
0\right\rangle _{1}\left| 0\right\rangle _{2}+\alpha _{1}\left|
0\right\rangle _{1}\left| 1\right\rangle _{2}+\alpha _{2}\left|
1\right\rangle _{1}\left| 0\right\rangle _{2}+\alpha _{3}\left|
1\right\rangle _{1}\left| 1\right\rangle _{2}.$ Second, statistics is
recovered by symmetrizing $\left| \Psi \left( t\right) \right\rangle
^{^{\prime }}$, namely by projecting it on $H_{t}$. We take this didactic
procedure seriously: particle statistics is interpreted as the result of 
{\em projection of the system state on a predetermined Hilbert subspace},
the one which satisfies the symmetry. This amounts to considering the
equation 
\begin{equation}
\forall t:S_{12}\left| \Psi \left( t\right) \right\rangle =\left| \Psi
\left( t\right) \right\rangle ,
\end{equation}
as a {\em constraint} applied to $\left| \Psi \left( t\right) \right\rangle $%
. When a particle statistics symmetry is an initial condition conserved as a
constant of motion, this constraint is redundant. However, the notion of
Connection will be related to particle statistics by means of a
counterfactual reasoning based on eq. (1). The idea is that $\left| \Psi
\left( t\right) \right\rangle $, symmetrical at time $t$, {\em could} be
pushed out of symmetry at time $t+dt$; but in this case eq. (1) would
``immediately'' project it on $H_{t}$ back again. Particle statistics would
operate like a watch-dog effect internal to the endosystem, or like
destructive and constructive interference, by killing the amplitudes of
those eigenstates of $\left| \Psi \left( t+dt\right) \right\rangle $ which
violate the symmetry, and reinforcing the other amplitudes through
re-normalization. This can also be seen as a continuous form of partial
state vector reduction on a symmetrical subspace.

To see why $\left| \Psi \left( t\right) \right\rangle $ could be ``pushed
out of symmetry'', we must consider the system defined in Section I and the $%
A_{rs}$ symmetrization projector. In a first step, $A_{rs}$ projection is
disregarded while parts $r$ and $s$ are assumed to be independent of each
other. An operation on part $r$ could well push the overall state $\left|
\Psi \left( t\right) \right\rangle $ out of symmetry, but in a second step
this is prevented by the continuous projection of $\left| \Psi \left(
t\right) \right\rangle $ on $H_{s}$: $\forall t:A_{rs}\left| \Psi \left(
t\right) \right\rangle =\left| \Psi \left( t\right) \right\rangle .$

We should note that this projection (or, if one prefers, partial state
vector reduction on a predetermined subspace) will in general alter the
entanglement between the parts $r$ and $s$, thus the coherence elements of $%
\rho _{r}\left( t\right) $ (part $r$ density matrix). However, it does not
alter the diagonal of $\rho _{r}\left( t\right) $; this is determined by the
operation performed on part $r$, namely it is a constraint that should be
satisfied by projection.

\section{Behaviour of the quantum mechanical Connection}

We go back to the Connection $r$,$s$ defined in Section I, and consider an
operation performed on just one qubit, say, $r.$ Let this be the continuous
rotation $\cos \varphi \left| 0\right\rangle _{r}\left\langle 0\right|
_{r}-\sin \varphi \left| 0\right\rangle _{r}\left\langle 1\right| _{r}+\sin
\varphi \left| 1\right\rangle _{r}\left\langle 0\right| _{r}+\cos \varphi
\left| 1\right\rangle _{r}\left\langle 1\right| _{r}$, with $\varphi =\omega
t$ and time $t$ ranging from $0$ to $\frac{\varphi _{F}}{\omega }$. We shall
examine the effect of applying $Q_{r}\left( \varphi \right) $ to qubit $r$,

\begin{equation}
\rho _{r}\left( t\right) =Q_{r}\left( \omega t\right) \rho _{r}\left(
0\right) Q_{r}^{\dagger }\left( \omega t\right) ,
\end{equation}
{\em under continuous} $A_{rs}$ {\em projection} of the overall state.

Let the Connection initial state be the ``symmetrical'' state (whose tensor
products satisfy symmetry $A_{rs}$):

\begin{equation}
\left| \Psi \left( 0\right) \right\rangle =\cos \vartheta \left|
0\right\rangle _{r}\left| 1\right\rangle _{s}+\sin \vartheta \left|
1\right\rangle _{r}\left| 0\right\rangle _{s}.
\end{equation}
The successive states are obtained by submitting a free normalized vector $%
\left| \Psi \left( t\right) \right\rangle $ of the Hilbert space $H_{w}$
(Section I) to the following {\em mathematically simultaneous} conditions,

\noindent {\em for all t or} $\varphi $:

\begin{enumerate}
\item[i)]  $A_{rs}\left| \Psi \left( t\right) \right\rangle =\left| \Psi
\left( t\right) \right\rangle $;

\item[ii)]  $\rho _{r}\left( t\right) =Tr_{s}\left( \left| \Psi \left(
t\right) \right\rangle \left\langle \Psi \left( t\right) \right| \right)
=\cos ^{2}\left( \vartheta +\varphi \right) \left| 0\right\rangle
_{r}\left\langle 0\right| _{r}+\sin ^{2}\left( \vartheta +\varphi \right)
\left| 1\right\rangle _{r}\left\langle 1\right| _{r}$; $Tr_{s}$ means
partial trace over $s$. Under condition (i), $\left| \Psi \left( t\right)
\right\rangle $ has always the form $\alpha \left( t\right) \left|
0\right\rangle _{r}\left| 1\right\rangle _{s}+\beta \left( t\right) \left|
1\right\rangle _{r}\left| 0\right\rangle _{s}$, therefore $\rho _{r}\left(
t\right) $ is a diagonal matrix: its coherent elements have been killed by $%
A_{rs}$ projection, or reduction;

\item[iii)]  the distance between the vectors before and after projection
must be minimum. In the case of continuous projection, $\left\| \left\langle
\left. \Psi \left( t\right) \right| \Psi \left( t+\triangle t\right)
\right\rangle \right\| $ must be maximized orderly for $t=0$, $t=\triangle t$%
, $t=2\triangle t$, $...$, $t=N\triangle t$, where $\triangle t=\frac{%
\varphi _{F}}{N\omega }$; then the limit for $N\rightarrow \infty $ should
be taken (however, maximization ordering turns out to be irrelevant here).
\end{enumerate}

\noindent (i) and (ii) yield $\left| \Psi \left( t\right) \right\rangle
=\cos \left( \vartheta +\varphi \right) \left| 0\right\rangle _{r}\left|
1\right\rangle _{s}+e^{i\delta }\sin \left( \vartheta +\varphi \right)
\left| 1\right\rangle _{r}\left| 0\right\rangle _{s},$ with $\delta $
unconstrained, as can be checked; condition (iii), given the initial state
(3), sets $\delta =0$, yielding to the unitary evolution:

\begin{equation}
\left| \Psi \left( t\right) \right\rangle =\cos \left( \vartheta +\varphi
\right) \left| 0\right\rangle _{r}\left| 1\right\rangle _{s}+\sin \left(
\vartheta +\varphi \right) \left| 1\right\rangle _{r}\left| 0\right\rangle
_{s}.
\end{equation}

This makes a ``good'' Connection. The rotation of qubit $r$ is identically
transmitted to the other qubit $s$. In fact

\begin{equation}
Tr_{r}\left( \left| \Psi \left( t\right) \right\rangle \left\langle \Psi
\left( t\right) \right| \right) =\rho _{s}\left( t\right) =\sin ^{2}\left(
\vartheta +\varphi \right) \left| 0\right\rangle _{s}\left\langle 0\right|
_{s}+\cos ^{2}\left( \vartheta +\varphi \right) \left| 1\right\rangle
_{s}\left\langle 1\right| _{s}.
\end{equation}
Of course eigenvalues 0 and 1 are interchanged: one qubit is the NOT\ of the
other. Noticeably, by simultaneously rotating also the other extremity $s$
of the Connection by the same amount, the same result (4) is obtained. This
means adding eq. (5) as a condition, but this is redundant with respect to
(i) and (ii), it was derived from (i) and (ii). Whereas, two different
rotations of the two Connection extremities give an impossible mathematical
system; this resembles a rigid classical Connection.

It should be noted that a rotation $\varphi $ of qubit (part) $r$ under $%
A_{rs}$ projection, is equivalent to applying the unitary operator $Q\left(
\varphi \right) $ to $\left| \Psi \left( t\right) \right\rangle $:

$Q\left( \varphi \right) \equiv \left( 
\begin{array}{cccc}
\cos \varphi  & \sin \varphi  & 0 & 0 \\ 
-\sin \varphi  & \cos \varphi  & 0 & 0 \\ 
0 & 0 & \cos \varphi  & -\sin \varphi  \\ 
0 & 0 & \sin \varphi  & \cos \varphi 
\end{array}
\right) $, with $\left| 0\right\rangle _{r}\left| 1\right\rangle _{s}\equiv
\left( 
\begin{array}{c}
1 \\ 
0 \\ 
0 \\ 
1
\end{array}
\right) ,\left| 1\right\rangle _{r}\left| 0\right\rangle _{s}\equiv \left( 
\begin{array}{c}
0 \\ 
1 \\ 
1 \\ 
0
\end{array}
\right) .$

\noindent $Q\left( \varphi \right) $ brings from $\left| \Psi \left(
0\right) \right\rangle $ (3) to $\left| \Psi \left( t\right) \right\rangle $
(4) without ever violating $A_{rs}$. We have thus ascertained a peculiar
fact. Our operation on a part, {\em blind} to its effect on the whole,
performed together with continuous $A_{rs}$ projection, generates a {\em %
unitary} {\em transformation} which is, so to speak, {\em wise} to the whole
state, to how it should be transformed without violating $A_{rs}$. Of course 
$A_{rs}$ ends up commuting with the resulting overall unitary propagator
(shaped by it).

\section{Quantum computation networks}

Let us consider the reversible Boolean network of fig. 2(a), fully deployed
in space $-$ time is orthogonal to the network lay-out. This is different
from sequential computation, where the Boolean network appears in the
computation space-time diagram.

Nodes $t$, $u$, $v$ and $r$ make the input and the output of a controlled
NOT; $r$ and $s$ belong to a Connection. This c-NOT is made up of four
coexisting qubits, and has four eigenstates which map the gate Boolean
relation and constitutes the basis of 
\[
H_{g}=span\{\left| 0\right\rangle _{t}\left| 0\right\rangle _{u}\left|
0\right\rangle _{v}\left| 0\right\rangle _{r},\left| 0\right\rangle
_{t}\left| 1\right\rangle _{u}\left| 0\right\rangle _{v}\left|
1\right\rangle _{r},\left| 1\right\rangle _{t}\left| 0\right\rangle
_{u}\left| 1\right\rangle _{v}\left| 1\right\rangle _{r},\left|
1\right\rangle _{t}\left| 1\right\rangle _{u}\left| 1\right\rangle
_{v}\left| 0\right\rangle _{r}\}.
\]
Model Hamiltonians of such gates are given in [1,2]; this is different from
time-sequential gates where the input and output are successive states of
the same register.$^{\left[ 3,4,5,6\right] }$

The satisfiability problem is stated by constraining {\em part} of the input
and {\em part} of the output, and asking whether this network admits a
solution. Let $u=1$ and $s=1$ be such constraints. $u=1$ ($s=1$) propagates
a {\em conditional} logical implication from left to right (right to left).
Logical implication is conditioned by the values of the unconstrained part
of the input (output). To have a solution, the two propagations must be
matched, i.e. they must generate a univocal set of values on all the nodes
of the network. Finding whether the network admits at least one match (one
solution) is an NP-complete problem. Possible collisions (mismatch) between
the two propagations will be both overcome and reconciled by the Connection.

Let us assume that the network has just one solution (which is the case
here: $t=1,$ $u=1,$ $r=0,$ $v=1,$ $s=1$). The procedure to find it is as
follows (this will hold for a generic network, thus we will think of many
gates and Connections $-$ in fig. 2b each wire is a Connection). The output
constraint is removed while an arbitrary value, here $t=0$, is assigned to
the unconstrained part of the input. The logical propagation of this input
toward the output yields $t=0$, $u=1$, $r=1$, $s=0$ ($v=t$ will be
disregarded). This computation is performed off line in polynomial time. It
serves to specify the initial state in which the network must be prepared: $%
\left| \Psi \left( 0\right) \right\rangle =\left| 0\right\rangle _{t}\left|
1\right\rangle _{u}\left| 1\right\rangle _{r}\left| 0\right\rangle _{s}$.
This state satisfies the gate/s and the Connection/s, but qubit $s$ is in $%
\left| 0\right\rangle _{s}$ $\left\langle 0\right| _{s}$ rather than $\left|
1\right\rangle _{s}$ $\left\langle 1\right| _{s}$ (the output constraint).
It will be continuously rotated from $\left| 0\right\rangle _{s}\left\langle
0\right| _{s}$ to $\left| 1\right\rangle _{s}\left\langle 1\right| _{s}$ 
{\em under} $A_{rs}$ {\em projection}, while keeping $\rho _{u}=\left|
1\right\rangle _{u}\left\langle 1\right| _{u}$ fixed. This transformation
operates on the network Hilbert space $H_{n}$; here $H_{n}=H_{g}\otimes
H_{s} $ where $H_{s}=span\left\{ \left| 0\right\rangle _{s},\left|
1\right\rangle _{s}\right\} $. Note that all states of $H_{n}$ natively
satisfy the gate/s, not necessarily the Connection/s.

At any time $t$, the state of the network is obtained by submitting a free
normalized state $\left| \Psi \left( t\right) \right\rangle $ of $H_{n}$ to
the conditions:

\noindent {\em for all t}:

\begin{enumerate}
\item[i)]  $A_{rs}\left| \Psi \left( t\right) \right\rangle =\left| \Psi
\left( t\right) \right\rangle $;

\item[ii)]  $Tr_{t,u,r}\left( \left| \Psi \left( t\right) \right\rangle
\left\langle \Psi \left( t\right) \right| \right) =\rho _{s}\left( t\right)
=\cos ^{2}\varphi \left| 0\right\rangle _{s}\left\langle 0\right| _{s}+\sin
^{2}\varphi \left| 1\right\rangle _{s}\left\langle 1\right| ,$ with $\varphi
=\omega t$ and $t$ going from $0$ to $\frac{\pi }{2\omega }$;

\item[iii)]  $Tr_{t,r,s}\left( \left| \Psi \left( t\right) \right\rangle
\left\langle \Psi \left( t\right) \right| \right) =\rho _{u}\left( 0\right)
=\left| 1\right\rangle _{u}\left\langle 1\right| _{u}$; in a generic network
there might be more conditions of this kind;

\item[iv)]  the distance between the vectors before and after projection
should be minimum, as specified in Section III.
\end{enumerate}

This yields: $\left| \Psi \left( t\right) \right\rangle =\cos \varphi \left|
0\right\rangle _{t}\left| 1\right\rangle _{u}\left| 1\right\rangle
_{r}\left| 0\right\rangle _{s}$ $+e^{i\delta }\sin \varphi \left|
1\right\rangle _{t}\left| 1\right\rangle _{u}\left| 0\right\rangle
_{r}\left| 1\right\rangle _{s}$, with $\varphi =\omega t$, as is readily
checked. Condition (iv) and the network initial state set $\delta =0$. Thus 
\begin{equation}
\left| \Psi \left( t\right) \right\rangle =\cos \varphi \left|
0\right\rangle _{t}\left| 1\right\rangle _{u}\left| 1\right\rangle
_{r}\left| 0\right\rangle _{s}+\sin \varphi \left| 1\right\rangle _{t}\left|
1\right\rangle _{u}\left| 0\right\rangle _{r}\left| 1\right\rangle _{s}
\end{equation}

For $\varphi =\frac{\pi }{2}$, one obtains $\left| \Psi \left( \frac{\pi }{%
2\omega }\right) \right\rangle =\left| 1\right\rangle _{t}\left|
1\right\rangle _{u}\left| 0\right\rangle _{r}\left| 1\right\rangle _{s}$,
namely the solution.

The unitary transformation (6) brings the state of the network from
satisfying only the input to satisfying both the input and the output
constraints. It is obtained by ``blindly'' operating on {\em divided parts}
of the network, but under $A_{rs}$ projection/s (the {\em conquering}
factor).

The evolution is always unitary because an infinitesimal rotation of $\rho
_{s}$, under conditions (i) through (iv), yields a univocal (unitary)
vector. We skip the lengthy but straightforward demonstration of this.

As a result of this process, $A_{rs}$ symmetries (or projectors) become
constants of motion which commute with the network propagator at all times.
They are also pairwise commuting, being applied to disjoint Hilbert spaces.
However, the cause should not be confused with the effect. $A_{rs}$
projections shape or forge the unitary propagator with which they commute.

If the network admits no solution, conditions (i) through (iv) make up an
impossible system. Measuring the network final state $-$ at $t=\frac{\pi }{%
2\omega }$ $-$ gives a non-solution. This is checkable in polynomial time
and tells that the network is not satisfiable.

If the network admits many solutions, the final state can be a linear
combination thereof. Which one, depends on the network initial state through
condition (iv). However measurement gives one solution (that it is a
solution is checkable in polynomial time).

It is clear from the above that Connections ``cut'' network complexity,
inducing a divide-and-conquer strategy. This diakoptic approach would make
NP-complete $\equiv $ P. However, we have applied reverse engineering until
now. The $A_{rs}$ projections are just a nice-to-have feature. This raises
the problem whether this feature is physical.

\section{Induced symmetry}

$A_{rs}$ symmetry will be shown to be an epiphenomenon of fermionic
antisymmetry in a special physical situation. This is generated by
submitting a couple of identical fermions 1 and 2 to a suitable Hamiltonian$%
^{[12]}$. We assume that each fermion has two compatible, binary degrees of
freedom $\chi $ and $\lambda $. Just for the sake of visualization (things
can remain more abstract), we can think that each fermion is a spin $1/2$
particle which can occupy one of either two sites of a spatial lattice. $%
\chi $ thus becomes the particle spin component $\sigma _{z}$ $\left( \chi
=0,1\text{ correspond to }\sigma _{z}=\text{{\em down, up}}\right) $ and $%
\lambda =r,s$ the label of the site occupied by the particle. For example, $%
\left| 0\right\rangle _{1}\left| 1\right\rangle _{2}\left| r\right\rangle
_{1}\left| s\right\rangle _{2}$ reads: $\sigma _{z}$ of particle 1 down (0), 
$\sigma _{z}$ of particle 2 up (1), site of particle $1\equiv r,$ site of
particle $2\equiv s$.

The following is the list of the states which do not violate statistics;
they make up the basis of the Hilbert space $H_{\lambda \chi }$. States are
represented in first and second quantization and, when there is exactly one
particle per site, in qubit notation (where $\sigma _{z}/\lambda $ are the
qubit eigenvalue/label):

$\left| a\right\rangle =\frac{1}{\sqrt{2}}\left( \left| 0\right\rangle
_{1}\left| 1\right\rangle _{2}-\left| 1\right\rangle _{1}\left|
0\right\rangle _{2}\right) \left| r\right\rangle _{1}\left| r\right\rangle
_{2}=a_{0r}^{\dagger }\ a_{1r}^{\dagger }\left| 0\right\rangle ,$

$\left| b\right\rangle =\frac{1}{\sqrt{2}}\left( \left| 0\right\rangle
_{1}\left| 1\right\rangle _{2}-\left| 1\right\rangle _{1}\left|
0\right\rangle _{2}\right) \left| s\right\rangle _{1}\left| s\right\rangle
_{2}=a_{0s}^{\dagger }\ a_{1s}^{\dagger }\left| 0\right\rangle ;$

$\left| c\right\rangle =\frac{1}{\sqrt{2}}\left| 0\right\rangle _{1}\left|
0\right\rangle _{2}\left( \left| r\right\rangle _{1}\left| s\right\rangle
_{2}-\left| s\right\rangle _{1}\left| r\right\rangle _{2}\right)
=a_{0r}^{\dagger }\ a_{0s}^{\dagger }\left| 0\right\rangle =\left|
0\right\rangle _{r}\left| 0\right\rangle _{s},$

$\left| d\right\rangle =\frac{1}{\sqrt{2}}\left| 1\right\rangle _{1}\left|
1\right\rangle _{2}\left( \left| r\right\rangle _{1}\left| s\right\rangle
_{2}-\left| s\right\rangle _{1}\left| r\right\rangle _{2}\right)
=a_{1r}^{\dagger }\ a_{1s}^{\dagger }\left| 0\right\rangle =\left|
1\right\rangle _{r}\left| 1\right\rangle _{s},$

$\left| e\right\rangle =\frac{1}{2}\left( \left| 0\right\rangle _{1}\left|
1\right\rangle _{2}+\left| 1\right\rangle _{1}\left| 0\right\rangle
_{2}\right) \left( \left| r\right\rangle _{1}\left| s\right\rangle
_{2}-\left| s\right\rangle _{1}\left| r\right\rangle _{2}\right) =$

$\frac{1}{\sqrt{2}}\left( a_{0r}^{\dagger }\ a_{1s}^{\dagger
}+a_{1r}^{\dagger }\ a_{0s}^{\dagger }\right) \left| 0\right\rangle =\frac{1%
}{\sqrt{2}}\left( \left| 0\right\rangle _{r}\left| 1\right\rangle
_{s}+\left| 1\right\rangle _{r}\left| 0\right\rangle _{s}\right) .$

$\left| f\right\rangle =\frac{1}{2}\left( \left| 0\right\rangle _{1}\left|
1\right\rangle _{2}-\left| 1\right\rangle _{1}\left| 0\right\rangle
_{2}\right) \left( \left| r\right\rangle _{1}\left| s\right\rangle
_{2}+\left| s\right\rangle _{1}\left| r\right\rangle _{2}\right) =$

$\frac{1}{\sqrt{2}}\left( a_{0r}^{\dagger }\ a_{1s}^{\dagger
}-a_{1r}^{\dagger }\ a_{0s}^{\dagger }\right) \left| 0\right\rangle =\frac{1%
}{\sqrt{2}}\left( \left| 0\right\rangle _{r}\left| 1\right\rangle
_{s}-\left| 1\right\rangle _{r}\left| 0\right\rangle _{s}\right) .$

\medskip \noindent Creation/annihilation operators form the algebra: $%
\left\{ a_{i}^{\dagger },a_{j}^{\dagger }\right\} =\left\{
a_{i},a_{j}\right\} =0,\ \ \left\{ a_{i}^{\dagger },a_{j}\right\} =\delta
_{i,j}.$ Now we introduce the Hamiltonian $H_{rs}=E_{a}\left| a\right\rangle
\left\langle a\right| +E_{b}\left| b\right\rangle \left\langle b\right| +$ $%
E_{c}\left| c\right\rangle \left\langle c\right| +E_{d}\left| d\right\rangle
\left\langle d\right| $ or, in second quantization

\begin{eqnarray*}
H_{rs} &=&-(E_{a}\ a_{0r}^{\dagger }\ a_{1r}^{\dagger }a_{0r}a_{1r}+E_{b}\
a_{0s}^{\dagger }\ a_{1s}^{\dagger }a_{0s}a_{1s} \\
&&+E_{c}\ a_{0r}^{\dagger }\ a_{0s}^{\dagger }a_{0r}a_{0s}+E_{d}\
a_{1r}^{\dagger }\ a_{1s}^{\dagger }a_{1r}a_{1s}),
\end{eqnarray*}

\noindent with $E_{a}$, $E_{b}>E_{c}$, $E_{d}\geq E$ discretely above 0.
This leaves us with two degenerate ground eigenstates:

\[
\left| e\right\rangle =\frac{1}{\sqrt{2}}\left( \left| 0\right\rangle
_{r}\left| 1\right\rangle _{s}+\left| 1\right\rangle _{r}\left|
0\right\rangle _{s}\right) \text{ and }\left| f\right\rangle =\frac{1}{\sqrt{%
2}}\left( \left| 0\right\rangle _{r}\left| 1\right\rangle _{s}-\left|
1\right\rangle _{r}\left| 0\right\rangle _{s}\right) . 
\]
Alternatively, their linear combinations $\left| 0\right\rangle _{r}\left|
1\right\rangle _{s}$ and $\left| 1\right\rangle _{r}\left| 0\right\rangle
_{s}$ can be used as the two orthogonal ground eigenstates. The generic
ground state is thus:

\begin{equation}
\left| \Psi \right\rangle =\alpha \left| 0\right\rangle _{r}\left|
1\right\rangle _{s}+\beta \left| 1\right\rangle _{r}\left| 0\right\rangle
_{s}\text{, with }\left| \alpha \right| ^{2}+\left| \beta \right| ^{2}=1,
\end{equation}
which satisfies $A_{rs}$ symmetry.

Let $A_{12}\left| \Psi \right\rangle =\frac{1}{2}$ $\left( 1-P_{12}\right) $
be the antisymmetrization projector. Due to the anticommutation relations: $%
A_{12}\left| 0\right\rangle _{r}\left| 1\right\rangle _{s}=\left|
0\right\rangle _{r}\left| 1\right\rangle _{s}$ and $A_{12}\left|
1\right\rangle _{r}\left| 0\right\rangle _{s}=\left| 1\right\rangle
_{r}\left| 0\right\rangle _{s}$. Also, $A_{12}\left| 0\right\rangle
_{r}\left| 0\right\rangle _{s}=\left| 0\right\rangle _{r}\left|
0\right\rangle _{s}$ and $A_{12}\left| 1\right\rangle _{r}\left|
1\right\rangle _{s}=\left| 1\right\rangle _{r}\left| 1\right\rangle _{s}$,
without forgetting that these are excited states.

The Connection can be implemented by suitably operating on the ground state
(7). We assume that the initial (``symmetrical'') state of the Connection is
given by eq. (3): $\left| \Psi \left( 0\right) \right\rangle =\cos \vartheta
\left| 0\right\rangle _{r}\left| 1\right\rangle _{s}+\sin \vartheta \left|
1\right\rangle _{r}\left| 0\right\rangle _{s}$. Then transformation (2) $%
\left[ \rho _{r}\left( t\right) =Q_{r}\left( \omega t\right) \rho _{r}\left(
0\right) Q_{r}^{\dagger }\left( \omega t\right) \right] $ is applied to
qubit $r$, under continuous $A_{rs}$ projection. Let $\left| \Psi \left(
t\right) \right\rangle $ be a free normalized vector of $H_{\lambda \chi }$.
The Connection state at time $t$ is obtained by submitting $\left| \Psi
\left( t\right) \right\rangle $ to the following mathematically simultaneous
conditions, \newline
\smallskip {\em for all }$t$:

\begin{enumerate}
\item[i)]  $A_{12}\left| \Psi \left( t\right) \right\rangle =\left| \Psi
\left( t\right) \right\rangle $;
\end{enumerate}

\smallskip

\begin{enumerate}
\item[ii)]  $\rho _{r}\left( t\right) =Tr_{s}\left( \left| \Psi \left(
t\right) \right\rangle \left\langle \Psi \left( t\right) \right| \right)
=\cos ^{2}\left( \vartheta +\varphi \right) \left| 0\right\rangle
_{r}\left\langle 0\right| _{r}+\sin ^{2}\left( \vartheta +\varphi \right)
\left| 1\right\rangle _{r}\left\langle 1\right| _{r}$;
\end{enumerate}

\smallskip

\begin{enumerate}
\item[iii)]  the distance between the vectors before and after reduction
must be minimum, as specified in Section III;
\end{enumerate}

\smallskip

\begin{enumerate}
\item[iv)]  the expected Connection energy: $\left\langle \xi \left(
t\right) \right\rangle =\left\langle \Psi \left( t\right) \right|
H_{rs}\left| \Psi \left( t\right) \right\rangle ,$ must be minimum. Since
this minimum will always be {\em zero}, time ordering is irrelevant.
\end{enumerate}

\smallskip

\noindent It is readily seen that the solution of this system is still $%
\left| \Psi \left( t\right) \right\rangle $ of eq. (4): 
\[
\left| \Psi \left( t\right) \right\rangle =\cos \left( \vartheta +\varphi
\right) \left| 0\right\rangle _{r}\left| 1\right\rangle _{s}+\sin \left(
\vartheta +\varphi \right) \left| 1\right\rangle _{r}\left| 0\right\rangle
_{s}.
\]
Simultaneous satisfaction of (i), i.e. fermionic antisymmetry seen as
projection, and (iv) (which is satisfied by $\left\langle \xi \left(
t\right) \right\rangle =0$) originates the projection $A_{rs}\left| \Psi
\left( t\right) \right\rangle =\left| \Psi \left( t\right) \right\rangle $,
as is readily seen. Therefore, if $\left\langle \xi \left( t\right)
\right\rangle =0$, namely if the operation on qubit $r$ is performed {\em %
adiabatically}, we obtain the Connection.

Since this computation is {\em reversible}$^{\left[ 7,8\right] }${\em , }%
namely it does not dissipate free energy (the result of {\em driving and
shaping} is a unitary evolution), in principle $\left\langle \xi \left(
t\right) \right\rangle $ can always be zero. This is of course an
idealization, for the time being we are highlighting a possible, speculative
way of dealing with NP-complete problems.

By the way we should note that the tensor products $\left| 0\right\rangle
_{r}\left| 0\right\rangle _{s}$ and $\left| 1\right\rangle _{r}\left|
1\right\rangle _{s}$ that would be projected off since they violate particle
statistics (Section II), are {\em not} the antisymmetrical excited states $%
\left| c\right\rangle $ and $\left| d\right\rangle $, which satisfy $A_{12}$%
. They would be instead the symmetrical states 
\begin{eqnarray*}
\left| 0\right\rangle _{r}\left| 0\right\rangle _{s} &=&\frac{1}{\sqrt{2}}%
\left| 0\right\rangle _{1}\left| 0\right\rangle _{2}\left( \left|
r\right\rangle _{1}\left| s\right\rangle _{2}+\left| s\right\rangle
_{1}\left| r\right\rangle _{2}\right)  \\
\left| 1\right\rangle _{r}\left| 1\right\rangle _{s} &=&\frac{1}{\sqrt{2}}%
\left| 1\right\rangle _{1}\left| 1\right\rangle _{2}\left( \left|
r\right\rangle _{1}\left| s\right\rangle _{2}+\left| s\right\rangle
_{1}\left| r\right\rangle _{2}\right) 
\end{eqnarray*}
The two kinds of states (antisymmetrical and symmetrical) have the same
qubit notations and density matrices. We are of course in counterfactual
reasoning; the important thing is that conditions (i) through (iv) yield the
solution (4).

Let us now address the problem of creating many Connections, namely an $%
H_{rs}$ Hamiltonian per network wire $r,s$ (fig. 2a). These $H_{rs}$ operate
on disjoint pairs of qubits. Viewed as $A_{rs}$ projectors (which is the
case when $\left\langle \xi \left( t\right) \right\rangle =0$), they are
pairwise commuting. Still in the idealized case of adiabatic operation, the
Connections operate independently of each other.

\section{Conclusion}

The notion of applying a particle statistics symmetry (or projection) to
divide the quantum whole into parts without clipping its richness $-$ here
computation speed-up$^{\left[ 9,10,11,\text{ among others}\right] }$ $-$
introduces an engineering (diakoptic) perspective in the design of quantum
mechanisms. For the time being, the development of this idea remains at an
abstract level. Finding model Hamiltonians which implement the Hermitean
matrix of Section V could possibly be the next step.

The interpretation of particle statistics symmetry as projection on a
predetermined subspace is best modeled in a two-way (advanced and retarded
in time) propagation scheme$^{\left[ 12,13,14\right] }$.

This research has been supported by Elsag Bailey a Finmeccanica company.
Thanks are due to A. Ekert, D. Finkelstein, L. Levitin, S. LLoyd, C.
Macchiavello and T. Toffoli for useful suggestions.


\begin{references}
\bibitem{}  G. Castagnoli, {\em Int. J. Mod. Phys. B, }{\bf 13}, 2253 (1991).

\bibitem{}  G. Castagnoli and M. Rasetti, {\em Int. J. Theor. Phys.}, {\bf 32%
}, 2335 (1993).

\bibitem{}  A. Barenco, D. Deutsch, A. Ekert, R. Jozsa, {\em Phys. Rev. Lett.%
} {\bf 7}4, 4083 (1995).

\bibitem{}  D.P. Di Vincenzo, {\em Phys. Rev. A} {\bf 50}, 1015 (1995).

\bibitem{}  A. Barenco, D. Deutsch, A. Ekert, {\em Proc. R. Soc. London A} 
{\bf 449}, 669 (1995).

\bibitem{}  S. LLoyd, ${\em Phys.Rev.Lett.}$ {\bf 75}, 346 (1995).

\bibitem{}  C.H. Bennett, ``Logical Reversibility of Computation'' {\em IBM
J. Res. Dev.} {\bf 6}, 525 (1979).

\bibitem{}  E. Fredkin and T. Toffoli, {\em Int. J. Theor. Phys.}{\it \ }%
{\bf 21}{\it ,} 219 (1982){\it .}

\bibitem{}  D. Deutsch and R. Jozsa, {\em Proc. Roy. Soc. London} {\bf A 439}%
, 553 (1992).

\bibitem{}  D.R. Simon, {\em Proceedings of the 35th Annual Symposium on the
Foundation of Computer Science}, Los Alamitos, CA, {\bf 116} (1994).

\bibitem{}  P.W. Shor, {\em Proceedings of the 35th Annual Symposium on the
Foundation of Computer Science}, Los Alamitos, CA, {\bf 124} (1994).

\bibitem{}  G. Castagnoli, ``{\em Quantum Computation Based on Retarded and
Advanced Propagation}'', Boston PhysComp 96. Available on the Web
(HTTP://xxx.lanl.gov): quant-ph/9706019, to be published in Physica D.

\bibitem{}  J.G. Cramer, {\em Rev. Mod. Phys.} {\bf 58}, 647 (1989).

\bibitem{}  G. Castagnoli, ``Merging quantum computation and particle
statistics'', to be published in {\em Int. J. Theor. Phys. }1998.01.
\end{references}
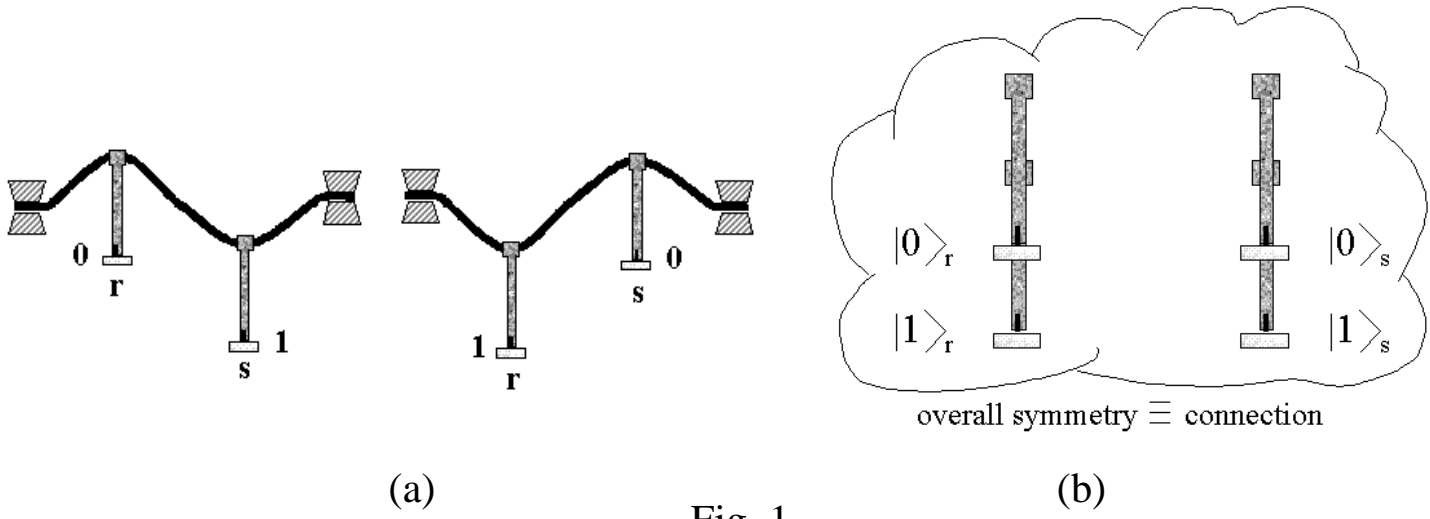
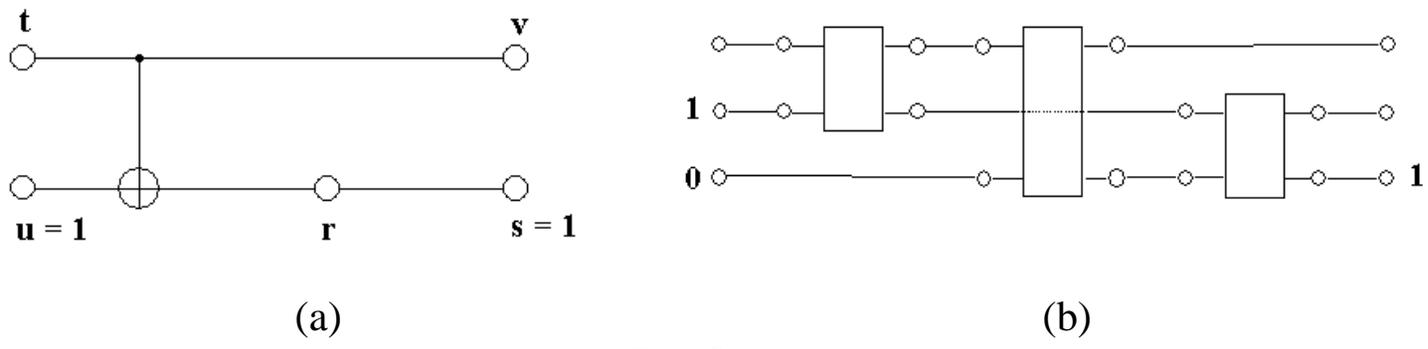
\end{document}